**Thermodynamic phase transition, pairing symmetry and Fermi surface topology in Ruddlesden-Popper nickelate films**

Yu Miao[1,2,3†], Zhiwei Wang[1,2,3†], Hongxu Sun[1,2,3†], Jianchang Shen[1,2,3], Runqing Luan[1,2,3], Zhipeng Ou[1,2,3], Xinru Yong[1,2,3], Zhenyu Wang[1,2,3], Tao Wu[1,2,3], Haoyu Hu[1], Junfeng He[1,2,3]*, Xianhui Chen[1,2,3]*

[1]*Department of Physics, University of Science and Technology of China, Hefei, Anhui 230026, China*

[2]*Hefei National Laboratory, University of Science and Technology of China, Hefei 230088, China*

[3]*Hefei National Research Center for Physical Sciences at the Microscale, University of Science and Technology of China, Hefei 230026, China*

†These authors contributed equally.

*Corresponding author: jfhe@ustc.edu.cn, chenxh@ustc.edu.cn

**Ruddlesden-Popper (RP) nickelates provide an uncharted territory to explore high-transition-temperature (high-$T_C$) superconductivity and superconducting mechanism. Here, we investigate the electronic structure of a new type of high-$T_C$ superconducting RP nickelate heterostructure $La_2PrNi_2O_7/NdAlO_3$ by angle-resolved photoemission spectroscopy. A superconducting state is observed without a pseudogap state, enabling a direct measurement of the superconducting order parameter and a microscopic extraction of the electronic specific heat. The observed superconducting gap opens at $T_C$ with prominent coherence peaks, illustrating the emergence of nonzero order parameter upon entering the superconducting state. An electronic specific heat jump appears at $T_C$, further demonstrating a thermodynamic phase transition. The magnitude of the superconducting order parameter is quantified by the observed superconducting coherence peaks, and a nodeless behavior is unambiguously established in the absence of pseudogap. The underlying Fermi surface consists of α, β and γ pockets, exhibiting a multi-orbital nature. Strain**

**dependent measurements further reveal the γ pocket in all superconducting and non-superconducting films with different epitaxial strain. Our results establish the missing thermodynamic evidence for superconducting phase transition in nickelates. They also provide direct evidence for the symmetry of the superconducting order parameter and illustrate the relationship between Fermi surface topology and the emergence of superconductivity in RP nickelate films.**

In conventional superconductors, the appearance of a zero-resistance state with perfect diamagnetism is accompanied by the emergence of a nonzero superconducting order parameter and a jump in electronic specific heat that define the spontaneous symmetry breaking and thermodynamic phase transition [1,2]. Recently, high-$T_C$ superconductivity was discovered in nickelates, either under high pressure or in the form of epitaxially grown thin films [3-22]. Zero-resistance and perfect diamagnetism were reported by transport and magnetic measurements [3-22]. Both superconducting gap and pseudogap were observed in RP phase $(La,Pr,Sm)_3Ni_2O_7$ films grown on $SrLaAlO_4$ substrate [23]. However, the emergence of a nonzero superconducting order parameter at $T_C$ and the associated thermodynamic evidence for superconducting phase transition remain to be explored in nickelate superconductors. In this work, by investigating the electronic structure of a new type of RP phase high-$T_C$ superconducting film $La_2PrNi_2O_7/NdAlO_3$ [24], we observe a superconducting gap that opens at $T_C$ with prominent coherence peaks but in the absence of a pseudogap for the first time. As a result, we report three important findings: thermodynamic evidence for superconducting phase transition, a pure nodeless superconducting order parameter and the role of Fermi surface topology for superconductivity in RP nickelates.

As illustrated in Fig. 1a, $La_2PrNi_2O_7$ thin films are grown on $NdAlO_3$ substrate [24], where a compressive strain of ~2.14% is applied on the films, comparing to the bulk $La_3Ni_2O_7$. The resistance measurement shows a superconducting onset temperature ($T_C^{onset}$) of ~60 K and a zero-resistance temperature ($T_C^{zero}$) of ~33 K (Fig. 1b). By using ultra-high vacuum (UHV) cryogenic sample transfer technique [23], electronic structures of the thin film are measured by angle-resolved photoemission spectroscopy (ARPES). Multiple pockets are observed on the underlying Fermi surface (Fig. 1c). We follow the notations in previous reports [25] and label the underlying Fermi pockets and the corresponding energy bands as α, β and γ, hereafter. The γ pocket is clearly resolved on the underlying Fermi surface map (Fig. 1c). This is also evidenced by the momentum cut measured along the Brillouin zone (BZ) boundary, where both β and γ bands are observed near $E_F$ (Fig. 1d). These results unambiguously reveal the coexistence of multiple underlying Fermi pockets (α, β and γ) with both $dx^2$-$y^2$ and $dz^2$ orbitals in the superconducting $La_2PrNi_2O_7$/$NdAlO_3$ film (Figs. 1c-d).

A central debate in superconducting RP nickelates is whether the superconducting order parameter exhibits a node on the $dx^2$-$y^2$ band (α/β) along the BZ diagonal. Such information represents the key difference between d-wave and s-wave (s±) pairing symmetry in the material [18,20,26,27]. For this purpose, high resolution laser-based ARPES measurement is carried out along the BZ diagonal direction of the new $La_2PrNi_2O_7$/$NdAlO_3$ film (Fig. 1f). A finite superconducting gap is directly seen in the Fermi-Dirac divided photoelectron intensity plot (Fig. 1e, also see extended data Fig. 1). This superconducting gap is quantitatively presented by the symmetrized energy distribution curves (EDCs) near the Fermi momentum ($k_F$). Prominent superconducting coherence peaks are observed on the EDCs and a bogoliubov back-bending behavior is identified with the band top below the Fermi energy ($E_F$) (Fig. 1h). The magnitude of the superconducting gap is quantified by fitting the symmetrized EDC at $k_F$ (Fig. 1g, also see extended data Fig. 2), which yields a finite size of ~16 meV.

Temperature dependent measurements reveal a gradual reduction of the superconducting gap upon increasing temperature (Fig. 2a). Notably, the superconducting gap vanishes above the superconducting transition temperature ($T_C^{onset}$) defined by transport, demonstrating the absence of a pseudogap. This is quantitatively shown by the symmetrized EDC at $k_F$ as a function of temperature (Fig. 2c). The energy gap is completely closed above $T_C$, evidenced by the single EDC peak at $E_F$ (e.g. at 70 K and 90 K). In this context, our results establish a pure superconducting state without any possible competing order [28,29]. Therefore, the superconducting gap, quantified by the coherence peaks, represents the magnitude of the superconducting order parameter Δ. The temperature evolution of Δ is summarized in Fig. 2d, which becomes nonzero below $T_C$, indicating a spontaneous symmetry breaking for a thermodynamic phase transition.

In nickelates, the direct thermodynamic evidence for superconducting phase transition remains unknown, largely due to the technical challenges for specific heat measurements under high pressure or on thin films. Nevertheless, the observed emergence of a nonzero superconducting order parameter and the associated change of electron density of states provide a microscopic description of the electronic entropy and specific heat. As demonstrated in cuprates [2,30], the electronic specific heat coefficient $\gamma_s$ can be extracted by using the electron density of states (DOS) measured by ARPES, which yields the same result as that of the bulk specific heat measurement [2]. In such a practice, $\gamma_s = dS/dT$, $S = -k_B\int[f\ln f + (1-f)\ln(1-f)]/f \text{ DOS}(E)\, dE$, where $S$ is the electronic entropy, $k_B$ is the Boltzmann constant, $f$ is the Fermi function, and DOS($E$) is achieved by the momentum integrated spectral weight measured by ARPES [2]. It is clear that $\gamma_s$ primarily scales with the temperature derivative of DOS($E$) near $E_F$. Following the practice in cuprates, we obtain the DOS($E$) by integrating the photoemission spectra over momentum (Fig. 3a, also see extended data Fig. 3). A sudden drop of DOS($E$) at $E_F$ is discernible across $T_C$ (~60 K, lower inset of Fig. 3a). For a quantitative analysis, the DOS($E$) around $E_F$ is integrated and labeled as DOS($E_F$). Then, the normalized DOS($E_F$) is shown as a function of temperature [$\text{DOS}(E_F)^{\text{Nor}}{}_T$ in Fig. 3b]. A sudden drop of the electron density of states is directly identified across $T_C$, resulting in a peak at ~$T_C$ in the temperature derivative $d\text{DOS}(E_F)^{\text{Nor}}{}_T/dT$ (Fig. 3c). Concomitantly, the extracted electronic specific heat coefficient $\gamma_s^{\text{DOS}}$ exhibits a jump at ~$T_C$ (Fig. 3c). Alternatively, the electronic entropy $S$ and specific heat coefficient $\gamma_s$ can also be estimated by the measured superconducting order parameter $\Delta$ [1,2,31]. The resulting $S^{\Delta}$ is shown in Fig. 3d, and a specific heat jump is identified at ~$T_C$ in $\gamma_s^{\Delta}$ (Fig. 3e). We note that the relations in the above two types of analysis become inexact with strong electronic correlation. Nevertheless, as established in cuprates, a qualitative demonstration of the electronic specific heat jump remains valid, especially when the superconducting fluctuation is absent above $T_C$ in our system [2,30]. As such, our results provide the thermodynamic evidence for superconducting phase transition from direct measurements of the electron density of states.

Another key issue in RP nickelate films is the relationship between Fermi surface topology and the emergence of superconductivity. In particular, the role of $\gamma$ Fermi pocket ($dz^2$ orbital) remains under intense debate [18,20,22,23,32-36]. For this purpose, we fix the chemical contents and oxygenation conditions of the $La_2PrNi_2O_7$ thin film [24], but systematically change the epitaxial strain by using different substrates (Fig. 4). Surprisingly, the γ pocket and the associated dispersive γ bands are observed in all superconducting and non-superconducting thin films, with either compressive or tensile strain (Figs. 4a, d-g). This is distinct from the results in bulk $La_3Ni_2O_7$, where a flat γ band-top is observed well below $E_F$ [25]. The size of γ pocket shows a moderate change with different substrates, as illustrated by the distance between the two γ bands along the (-π, π)-(0, π)-(π, π) direction (Fig. 4d-h). On the contrary, the size of β pocket exhibits prominent changes with different substrates, evidenced by the substantially altered distance between the two β bands near (0, π) (Fig. 4d-h). This is also visualized by Fermi surface map with 103 eV photons, which selectively probes the β pocket due to ARPES matrix-element effects (Fig. 4i-l). The strain dependent evolution of the (underlying) Fermi surface is summarized in Fig. 4m. Comparing the bulk $La_3Ni_2O_7$, a substantial amount of hole doping is induced in our $La_2PrNi_2O_7$ thin films (Fig. 4m).

After describing the experimental observations, we now discuss possible implications. First, the observation of a superconducting gap without pseudogap enables the identification of a nonzero superconducting order parameter that emerges at $T_C$, pointing to a spontaneous symmetry breaking for the phase transition. The extracted electronic specific heat jump further establishes the missing thermodynamic evidence for the superconducting phase transition. These results directly reveal the formation of a thermodynamically stable superconducting phase with coherent condensates, thus providing the last piece of puzzle for establishing a nickelate high-$T$c superconductor. Second, the nodeless superconducting gap in the absence of a pseudogap unambiguously demonstrates that the symmetry of the superconducting order parameter in RP nickelate films is inconsistent with d-wave. Instead, s-wave (s±) is supported by our experiment. Third, the γ pocket ($dz^2$ orbital) is observed in all superconducting and non-superconducting $La_2PrNi_2O_7$ thin films, ruling out a one-to-one correspondence between the emergence of γ Fermi pocket and superconductivity. On the other hand, the γ pocket indeed exists on the underlying Fermi surface of our superconducting $La_2PrNi_2O_7/NdAlO_3$ and $La_2PrNi_2O_7/SrLaAlO_4$ thin films. Therefore, it would naturally participate and contribute to the superconductivity. A notable observation is the steep dispersion of γ band near $E_F$ in the thin films, which exhibits little change with different substrates. This is distinct from the bulk $La_3Ni_2O_7$, where a flat band-top is observed below $E_F$. Such an evolution can be qualitatively captured by first-principles calculations if we consider a substantial hole doping in the thin films (Fig. 4c, also see extended data Fig. 4). As such, $E_F$ of the thin films locates in a region where the calculated γ band shows a steep dispersion and exhibits little change with different strain (Fig. 4c). In the meantime, the calculated β band shifts substantially with different strain, which is also consistent with the experimental observations (Fig. 4m). The origin of hole doping in the thin films remains unclear, but interstitial oxygen [34,37,38] and isovalent doping from Pr are possible candidates. Finally, since the Fermi surface topology remains qualitatively insensitive to epitaxial strain, it would be instructive to consider other key ingredients that may change with strain and trigger the superconductivity. In this context, it has been proposed that the octahedral tilts may be suppressed with compressive strain [39], increasing the out-of-plane $dz^2$ orbital coupling [28,40]. Prominent electron-boson coupling is also observed in superconducting thin films (see extended data Fig. 1) [23]. It would be interesting to investigate whether the compressive strain may help stabilize the bosonic mode (e.g. lattice vibration) that facilitates the emergence of superconductivity.

To summarize, we reveal the emergence of a nodeless superconducting order parameter and an electronic specific heat jump at $T_C$ in a new type of RP nickelate film $La_2PrNi_2O_7/NdAlO_3$, thus providing thermodynamic evidence for superconducting phase transition and direct insight into the pairing symmetry. Our strain dependent measurements clarify the relationship among Fermi surface topology, epitaxial strain and the emergence of superconductivity in RP nickelate films.

## Methods

### Thin film growth

$La_2PrNi_2O_7$ thin films were grown on as-received $NdAlO_3$(001), $SrLaAlO_4$(001), $LaAlO_3$(001), and (La,Sr)(Al,Ta)$O_3$(001) substrates (MTI-Kejing) using our optimized energy-switchable, ozone-assisted atomic-layer-by-layer epitaxy technique. By alternately ablating $La_{0.67}Pr_{0.33}O_x$ and $NiO_x$ targets using different laser energies under in situ reflection high-energy electron diffraction (RHEED) monitoring, we achieved atomic-layer-by-layer growth of films with exceptional crystalline quality under highly oxidizing conditions.

For the perovskite substrates, the ablation sequence was (La,Pr)O–(La,Pr)O–$NiO_2$–[(La,Pr)O–(La,Pr)O–$NiO_2$–(La,Pr)O–$NiO_2$] × n. For the $K_2NiF_4$-type substrate, the sequence was (La,Pr)O–$NiO_2$–[(La,Pr)O–(La,Pr)O–$NiO_2$–(La,Pr)O–$NiO_2$] × n. The film stoichiometry was controlled by precisely adjusting the number of laser pulses required to deposit a single atomic layer. Typically, 110–120 pulses were used to ablate the $La_{0.67}Pr_{0.33}O_x$ and $NiO_x$ targets, providing a stoichiometric precision of better than 1%.

The thin film growth was carried out at 750 °C in a mixed atmosphere of purified ozone and oxygen. The total growth pressure was maintained at 10 Pa, with an ozone partial pressure of approximately 2.1 Pa. The laser fluences were set to 1.6 and 2.0 J cm$^{-2}$ for the $La_{0.67}Pr_{0.33}O_x$ and $NiO_x$ targets, respectively, and the laser repetition rate was 4 Hz. Following deposition, the samples were cooled at a rate of 60 °C min$^{-1}$ to below 160 °C before being transferred from the growth chamber to the ultra-high vacuum (UHV) cryogenic sample transfer chamber to prevent oxygen loss. Further details of the growth process can be found in ref. [24].

**ARPES measurements**

Laser-based ARPES measurements were carried out at University of Science and Technology of China with a photon energy of 6.994 eV and a base pressure of better than $5 \times 10^{-11}$ Torr. The energy resolution of the measurements was ~ 2.5 meV. The Fermi level was referenced by measuring an Au piece in electrical contact with the samples. Synchrotron-based ARPES measurements were carried out at BL03U of Shanghai Synchrotron Radiation Facility (SSRF) with a base pressure of ~$5 \times 10^{-11}$ Torr.

**Calculation**

Density functional theory (DFT) calculations were performed using Quantum ESPRESSO within the plane-wave pseudopotential framework [42-45]. The exchange-correlation energy was treated within the Perdew-Burke-Ernzerhof (PBE) generalized gradient approximation [46]. A kinetic-energy cutoff of 100 Ry was used, together with a 12×12×1 or 12×12×2 Monkhorst-Pack k-point mesh [47]. Structural relaxations were carried out at fixed cell shape and volume, based on the experimental structure, by relaxing only the internal atomic coordinates. The single-layer limit was considered, where each unit cell contains only one $La_3Ni_2O_7$ layer, corresponding to two Ni atoms per unit cell, and a vacuum region was introduced along the z direction to suppress interlayer couplings. The Wannierization was performed using Wannier90, with the $dx^2$-$y^2$ and $dz^2$ orbitals retained for each Ni atom in both cases [48,49].

**Extraction of electronic specific heat coefficient from the measured superconducting gap Δ**

In a weakly correlated system, the superconducting density of states can be written as

$$D_{\mathrm{s}}^{0}(E,T) = \begin{cases} D_{\mathrm{N}}(E_{\mathrm{F}})\dfrac{|E|}{\sqrt{E^2-\Delta^2(T)}}, & |E| > \Delta(T) \\ 0, & |E| < \Delta(T) \end{cases}$$

where $D_{\mathrm{N}}(E_{\mathrm{F}})$ is the normal-state density of states at the Fermi level, $\Delta(T)$ is the experimentally determined superconducting gap Δ [50]. The experimental energy resolution can be included by convolving the density of states with a normalized Gaussian response function

$$D_{\mathrm{s}}(E,T) = \int_{-\infty}^{+\infty} D_{\mathrm{s}}^{0}(E',T)G_{\sigma}(E-E')dE'$$

where $G_{\sigma}(E) = \frac{1}{\sqrt{2\pi}\sigma_E}\exp\left(-\frac{E^2}{2\sigma_E^2}\right)$, and $\sigma_E = \frac{\mathrm{R}}{2\sqrt{2\ln 2}}$ is determined by the experimental energy resolution R (2.5 meV).

With the experimentally measured superconducting gap Δ, the electronic entropy $S^{\Delta}$was calculated by

$$S^{\Delta} = -k_{\mathrm{B}}\int D_{\mathrm{s}}(E,T)\Big[f(E,T)\ln[f(E,T)] + \big(1-f(E,T)\big)\ln[1-f(E,T)]\Big]dE$$

, and the electronic specific heat coefficient $\gamma_s^{\Delta}$ was obtained by $\gamma_{\mathrm{s}}^{\Delta} = \frac{dS^{\Delta}}{dT}$.

**Data availability:** All data are available in the main text or the extended data.

**Acknowledgments:** We acknowledge the support by the Scientific Research Innovation Capability Support Project for Young Faculty (SRICSPYF-ZY2025069), the National Key Research and Development Program of China (No. 2024YFA1408103), the CAS Superconducting Research Project (under Grant No. SCZX-0101), the Quantum Science and Technology-National Science and Technology Major Project (2021ZD0302802), the HFNL Self-Deployed Project (ZB2025020200), and the research start-up fund of University of Science and Technology of China.

**Author contributions:** J.H. and X.C. conceived the experiments and supervised the project; Y.M., H.S., J.S. and R.L. performed ARPES measurements with help from Z.O. and X.Y.; Z.W. grew and characterized the thin films with help from H.S.; J.H., Y.M., J.S. and H.S. developed the UHV cryogenic sample transfer system; H.H. performed the calculations; J.H. and X.C. interpreted the results with key inputs from Y.M., H.S., J.S., Z.O., Z.-Y.W. and T.W.; J.H., X.C. and Y.M. wrote the manuscript with discussions and comments from all authors.

**Competing interests:** Authors declare that they have no competing interests.

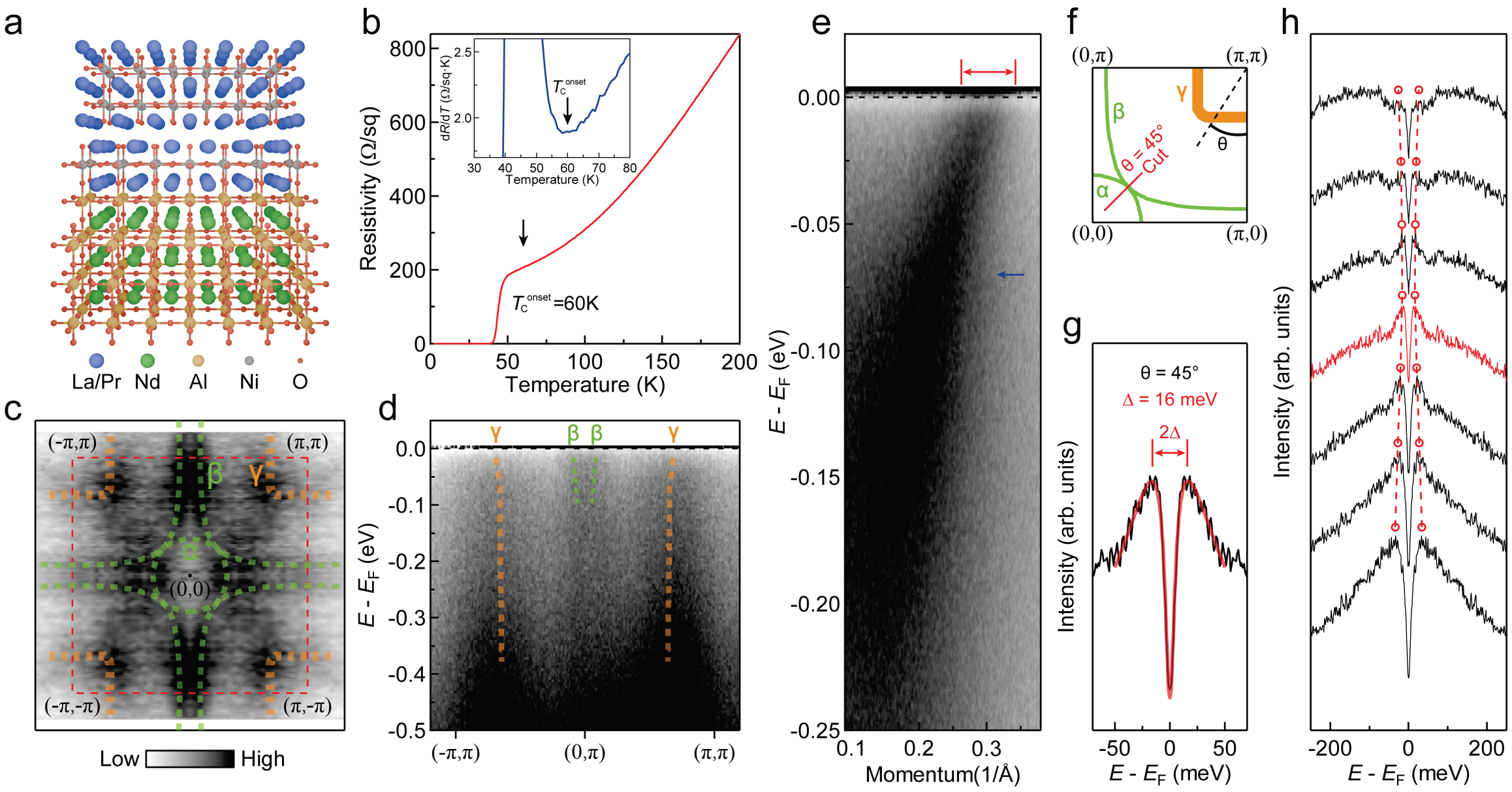

**Fig. 1. Electronic structure of $La_2PrNi_2O_7/NdAlO_3$.** (**a**) Crystal structure of $La_2PrNi_2O_7/NdAlO_3$. (**b**) Temperature dependent resistivity measurement of a 3-unit-cell (3UC) thick $La_2PrNi_2O_7/NdAlO_3$ thin film. The arrow marks the superconducting onset temperature $T_C^{onset}$, determined by the minimum value of $dR/dT$ in the inset. (**c**) Underlying Fermi surface map measured by synchrotron-based ARPES with 63 eV photons. (**d**) Fermi-Dirac-divided photoelectron intensity plot along the (-π,π)-(0,π)-(π,π) momentum path, measured by 63 eV photons. (**e**) Fermi-Dirac-divided photoelectron intensity plot along the BZ diagonal, measured by laser-based ARPES with 7 eV photons. (**f**) Schematic of the BZ. The red line marks the momentum cut in (e). (**g**) Symmetrized EDC at the Fermi momentum $k_F$. The red curve represents the fitting result by a phenomenological model used for high-$T_C$ superconductors (ref. 41). The fitting yields a superconducting gap of $\Delta = 16$ meV. (**h**) Symmetrized EDCs in a momentum region near $k_F$, marked by the red double-headed arrow in (e). The red curve corresponds to the EDC at $k_F$. All ARPES measurements were performed at 10 K.

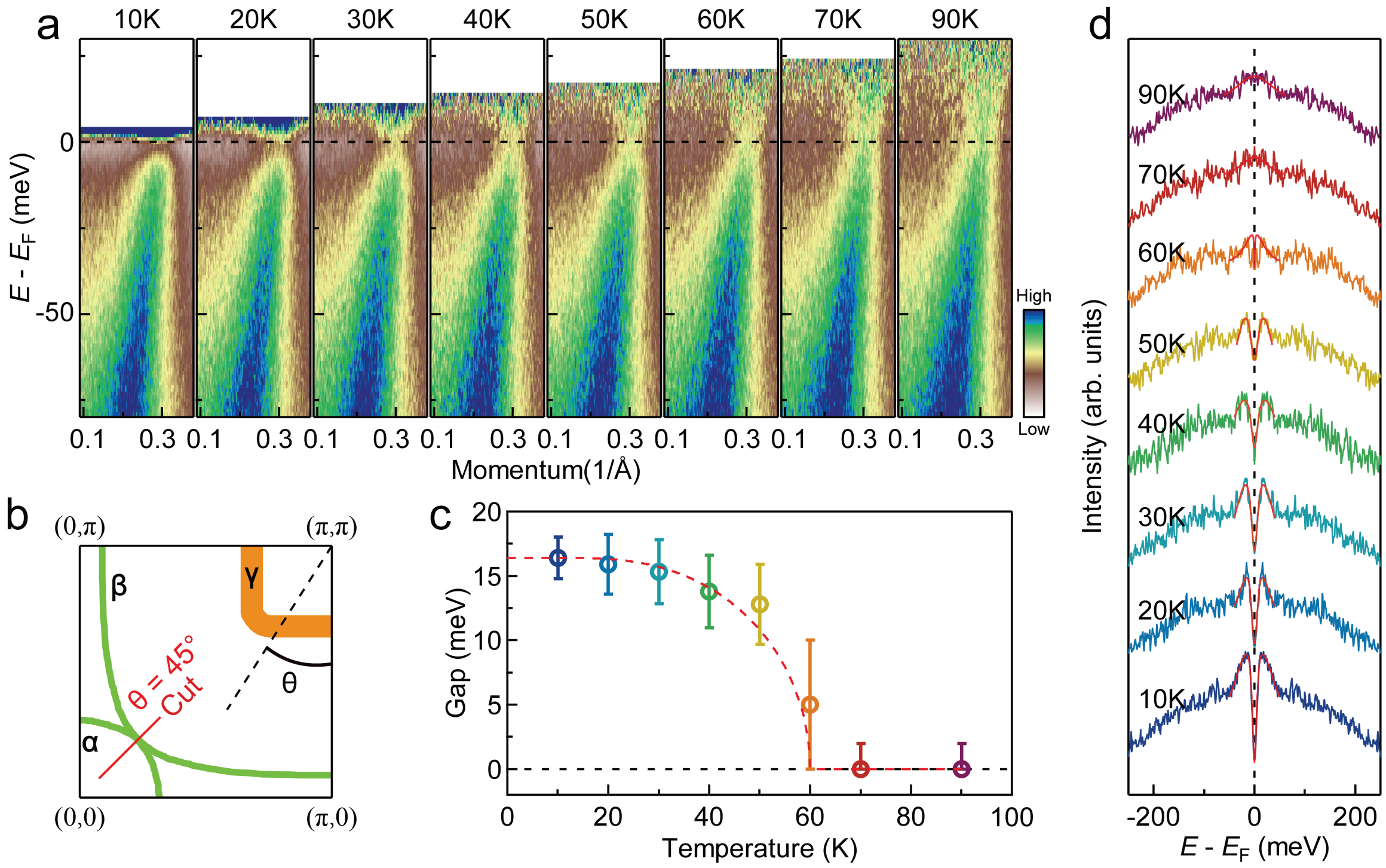

**Fig. 2. Temperature dependence of the superconducting gap in $La_2PrNi_2O_7/NdAlO_3$.** (**a**) Temperature-dependent photoelectron intensity plot measured along the BZ diagonal. Fermi-Dirac function is divided out to visualize the temperature evolution of the superconducting gap. (**b**) Schematic of the BZ and the location of the momentum cut in (a) (red line). (**c**) Temperature dependence of the extracted superconducting gap. Error bars represent the uncertainties in the determination of the gap magnitude. The dashed line shows a BCS-like temperature dependence. (**d**) Temperature dependence of the symmetrized EDC at $k_F$. Red curves are fits to a phenomenological spectral function (ref. 41).

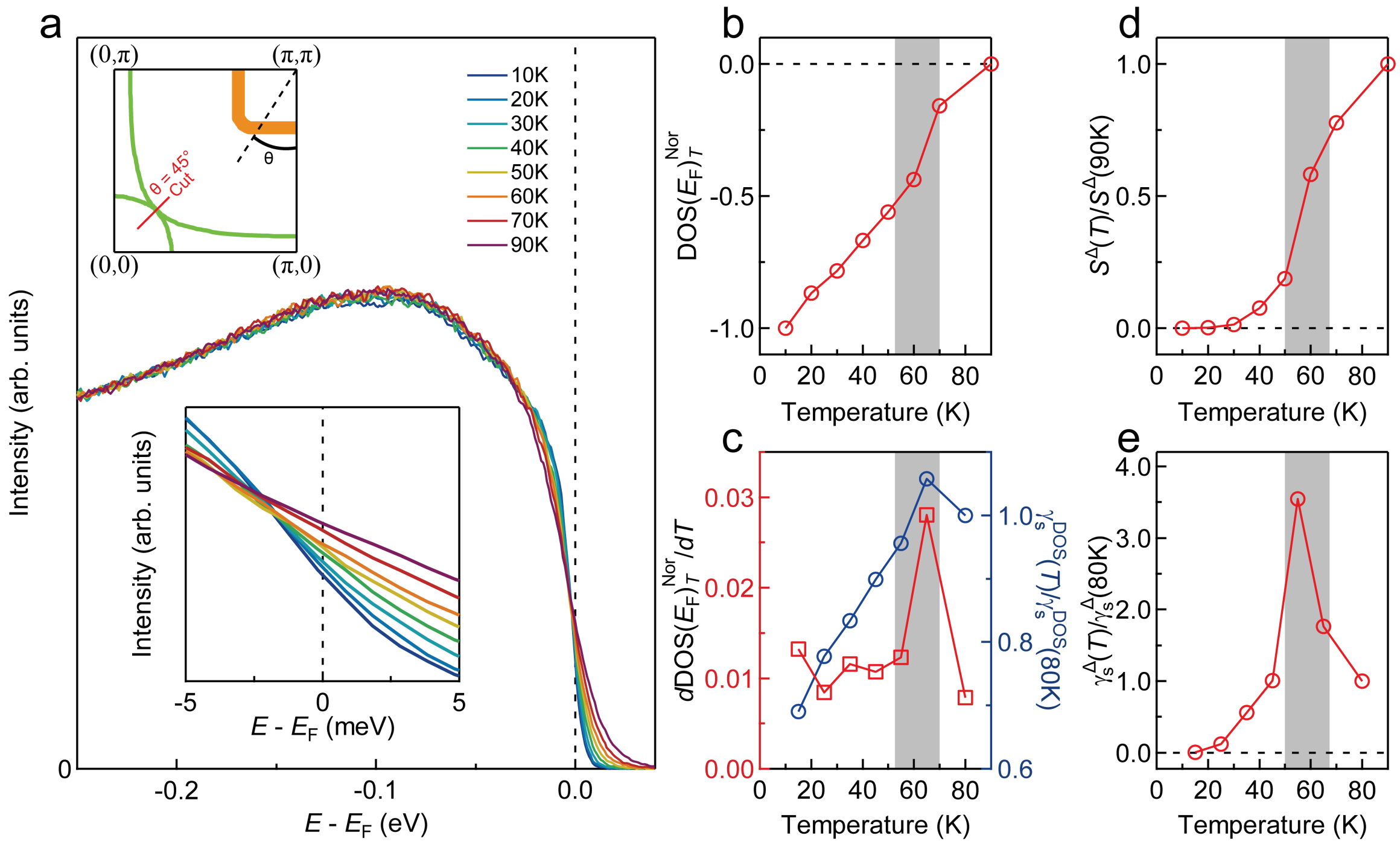

**Fig. 3. Temperature evolution of the electron density of states and electronic specific heat.** (**a**) Temperature evolution of the momentum-integrated EDC, measured along the BZ diagonal (red line in the schematic BZ). The energy region near $E_F$ is shown in the inset, illustrating a sudden drop of the electron DOS at $E_F$ below ~60 K. (**b**) Temperature dependence of the momentum-integrated spectral weight near $E_F$, labeled as DOS($E_F$), which is obtained by integrating the area of momentum-integrated EDC in the energy window of $E_F \pm 5$ meV. The normalized DOS($E_F$) is labeled as DOS$(E_F)^{Nor}{}_T$, which is defined by the relation DOS$(E_F)^{Nor}{}_T$ = [DOS$(E_F)_T$ - DOS$(E_F)_{90K}$] / ‖[DOS$(E_F)_{10K}$ - DOS$(E_F)_{90K}$]‖. (**c**) Temperature derivative of DOS$(E_F)^{Nor}{}_T$ (left axis), and electronic specific heat coefficient $\gamma_s^{DOS}$ extracted from the measured electron density of states (right axis). The $\gamma_s^{DOS}$ is normalized to its value at 80 K. (**d**) Temperature dependence of the electronic entropy $S^\Delta$ extracted from the measured superconducting gap. The $S^\Delta$ is normalized to its value at 90 K. (**e**) Temperature dependence of the electronic specific heat coefficient $\gamma_s^\Delta$ extracted from the measured superconducting gap. The $\gamma_s^\Delta$ is normalized to its value at 80 K. Gray shaded regions in (b-e) highlight the temperature where the electronic specific heat jump appears, being consistent with the $T_C^{onset}$ in transport measurements.

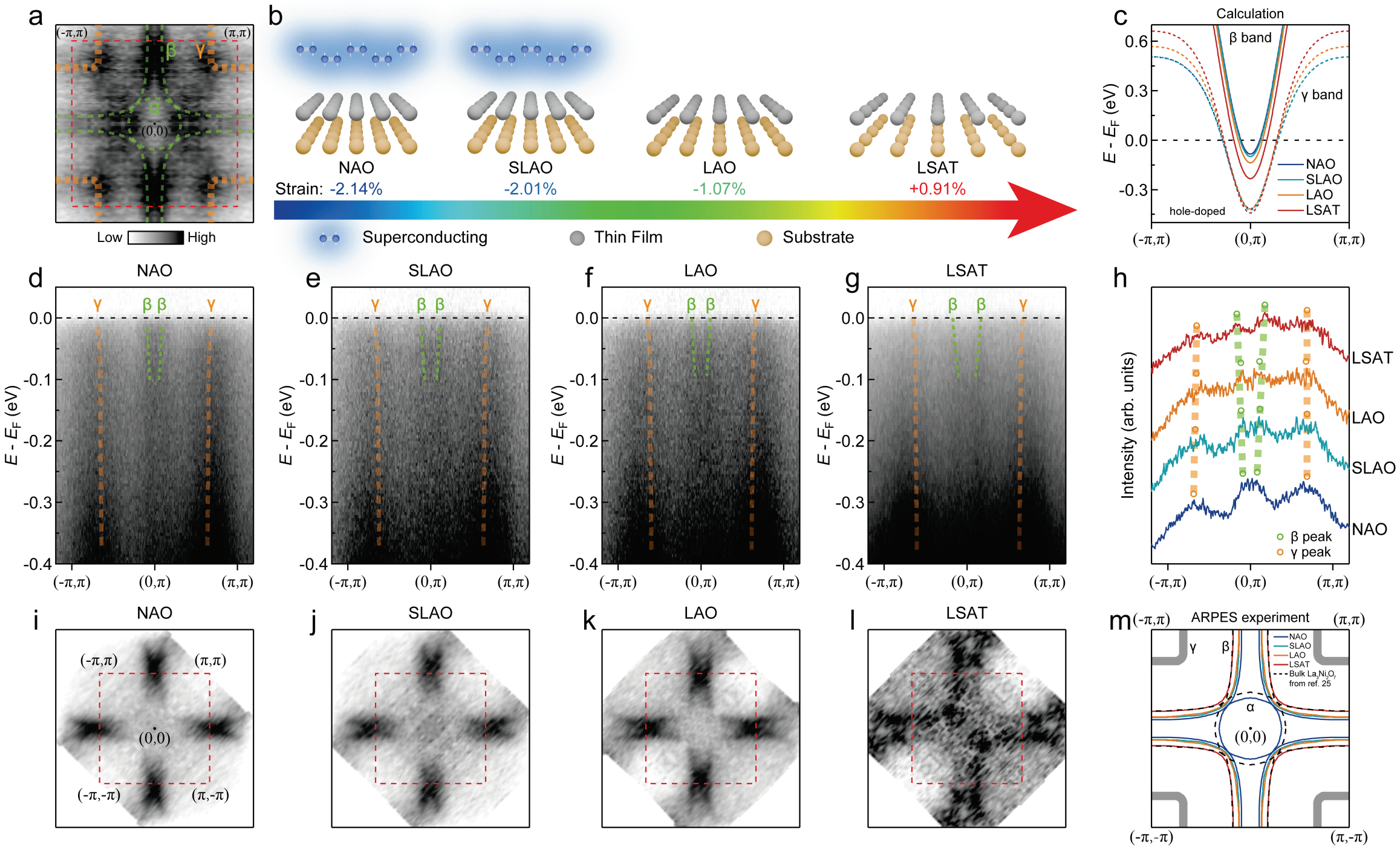

**Fig. 4. Evolution of Fermi surface topology with epitaxial strain in $La_2PrNi_2O_7$ thin films**. (**a**) Underlying Fermi surface intensity map of a superconducting $La_2PrNi_2O_7$ film grown on $NdAlO_3$ substrate, measured with 63 eV photons. (**b**) Schematic of the thin film with different epitaxial strain from the substrates. Superconductivity is observed above 10 K in the films with $NdAlO_3$ and $SrLaAlO_4$ substrates. (**c**) Calculated band dispersion along the (-π,π)-(0,π)-(π,π) direction of the $La_2PrNi_2O_7$ film grown on $NdAlO_3$ (NAO), $SrLaAlO_4$ (SLAO), $LaAlO_3$ (LAO) and $(La,Sr)(Al,Ta)O_3$ (LSAT) substrate, respectively. A substantial hole doping is included to mimic the experimental results. **(d-g)** Photoelectron intensity plots along the (-π,π)-(0,π)-(π,π) direction for the $La_2PrNi_2O_7$ film grown on NAO, SLAO, LAO and LSAT, respectively, measured with 63 eV photons. Green and orange dashed lines indicate the β and γ bands, respectively. (**h**) MDCs at $E_F$, extracted from (d-g), respectively. Green and orange circles mark the MDC peaks for the β and γ bands, respectively. The dashed lines are eye-guides for the substrate-dependent evolution of the peak positions. **(i-l)** Underlying Fermi surface intensity maps measured at 10 K, using 103 eV photons, which selectively enhance the β pocket by ARPES matrix-element effects. (**m**) Underlying Fermi surface extracted from the experiments, shown with the Fermi surface of bulk $La_3Ni_2O_7$ from ref. 25.